\documentclass[preprint,aps,showpacs,preprintnumbers,amsmath,amssymb,nofootinbib]
{revtex4}
\usepackage{epsfig}
\begin{document}

\begin{flushright}
\end{flushright}


\newcommand{\be}{\begin{equation}}
\newcommand{\ee}{\end{equation}}
\newcommand{\bea}{\begin{eqnarray}}
\newcommand{\eea}{\end{eqnarray}}
\newcommand{\nn}{\nonumber}

\def\lb{\Lambda_b}
\def\ll{\Lambda}
\def\mb{m_{\Lambda_b}}
\def\ml{m_\Lambda}
\def\s1{\hat s}
\def\ve{\varepsilon }
\def\ds{\displaystyle}


\title{\large Study of some rare decays of $B_s$ meson in the fourth generation model}
\author{R. Mohanta$^1$, A. K. Giri$^2$ }
\affiliation{
$^1$School of Physics, University of Hyderabad, Hyderabad - 500 046, India\\
$^2$Department of Physics, Indian Institute of Technology Hyderabad,
ODF Estate, Yedumailaram - 502205, Andhra Pradesh, India}

\begin{abstract}
We study some rare decays of $B_s$ meson governed by the quark
level transitions $ b \to s$,   in the fourth
generation model popularly known as SM4. Recently it has been shown that
SM4, which is a simple extension of the SM3, can successfully
explain several anomalies observed in the CP violation parameters
of $B$ and $B_s$ mesons. We find that in this model due to the
additional contributions coming from the heavy $t'$ quark in the loop,
the branching ratios and other observables in rare $B_s$
decays deviate significantly from their SM values. Some of these
modes are within the reach of LHCb experiment and search for such channels
are strongly argued.

\end{abstract}

\maketitle

\section{Introduction}

The spectacular performance of the two  asymmetric $B$ factories Belle and Babar
provided us an unique opportunity to understand the origin of CP violation in
a very precise way. Although,  the results from the $B$ factories
 do not provide us any clear evidence of
new physics, but there are few cases observed in the last few years,  which have
2-3 $\sigma$ deviations from their corresponding SM expectations \cite{hfag}. For example,
the difference between the direct CP asymmetry parameters between $B^- \to \pi^0
K^-$ and $\bar B^0 \to \pi^+ K^-$, which is expected to be negligibly small in the SM,
but found to be nearly $15 \%$. The measurement of mixing-induced CP asymmetry
in several $b \to s$ penguin decays  is not found to be same as that of
$B_d \to J/\psi K_s$. Recently, a very largish CP asymmetry has been
observed by the CDF and D0 collaborations \cite{cdf, d0} in the tagged analysis of
$B_s \to J/\psi \phi$  with value $S_{\psi \phi} \in [0.24,1.36]$.
Within the SM this asymmetry is expected to be vanishingly small,
which basically comes from $B_s - \bar B_s$ mixing phase. It should be
noted that all these deviations are associated with the flavour
changing neutral current (FCNC) transitions $b \to s$. It is well known that the
FCNC decays are forbidden at
the tree level in the standard model (SM) and therefore play a very
crucial role to look for the possible existence of  new physics (NP).

In this paper we would like to study some rare decays of $B_s$ meson involving
$b \to s$ transitions. The study of $B_s$ meson has attracted significant attention
in recent times because
huge number of $B_s$ mesons are expected to be produced in the currently running LHCb
experiment, which opened up the possibility to study $B_s$ meson with high statistical
precision. These studies will not only play a dominant role to corroborate the results
of $B$ mesons but also look for possible hints of new physics.
Here we will consider the decay channels  $B_s \to \phi \pi$,
 $B_s \to \phi \gamma $, $B_s \to \gamma \gamma$  and $B_s \to \mu^+ \mu^- \gamma$
which are highly suppressed in the SM. We intend to analyze these decay
channels both in the SM and in the fourth quark generation model \cite{4gen}, usually
known as SM4.
SM4 is a simple extension of the standard model with
three generations (SM3) with the additional up-type ($t'$) and
down-type ($d'$) quarks.
It has been shown in Ref. \cite{ref1}, that the addition of a fourth family of
quarks with $m_{t'}$ in the range
(400-600) GeV provides a simple explanation for the several deviations,
 that have been observed involving CP asymmetries in the
$B,~B_s$ decays. The implications of fourth generation in various
$B$ decays are discussed in \cite{4gnp, 4gnp1, rm3a, rm3,lenz}.
The experimental search for fourth generation quarks has also received
significant attention recently due to the operation of Large Hadron
Collider.
The CMS collaboration
put a lower bound on the mass of $t^\prime$ as  $m_{t^\prime}\gtrsim450\,$ GeV~\cite{cms1} and exclude
 the $b^\prime$-quark mass  in the region  255 GeV $<m_{b^\prime} <361 $ GeV at $95\%$ C.L.~\cite{cms2}.

The paper is organized as follows.In section II we discuss the non-leptonic decay
process $B_s \to \phi \pi$. The radiative decays $B_s \to \phi \gamma$ and $B_s \to \gamma
\gamma$ are discussed in Sections III and IV. The process $B_s \to
\mu^+ \mu^- \gamma $ is presented in  Section V and Section VI contains the Conclusion.

\section{$B_s \to \phi \pi$ Process}

In this section we will discuss the non-leptonic decay mode $B_s \to \phi \pi$
which receives dominant contribution from
 electroweak penguins $b \to s q \bar q$ ($q=u,d$), as the QCD penguins are OZI
suppressed, and the color-suppressed tree contribution $b \to u \bar u s$
is doubly Cabibbo suppressed. Therefore, this process is  expected to be highly suppressed in the
SM and hence serves as a suitable place to search for new
physics. This decay mode has been studied in the SM using QCD factorization approach
\cite{cheng} and in the model with non-universal $Z'$ boson \cite{kim}.

The relevant effective Hamiltonian describing this
process is given by \cite{rg}
\begin{equation}
{\cal H}_{eff}^{SM} = \frac{G_F}{\sqrt{2}} \left [ V_{ub}V_{us}^* \sum_{i=1,2} C_i(\mu) O_i
- V_{tb} V_{ts}^*
\sum_{i=3}^{10} C_i(\mu) O_i \right ]\;,
\end{equation}
where  $C_i(\mu)$'s are the Wilson coefficients evaluated
at the $b$-quark mass scale, $O_{1,2}$ are the tree level current-current operators,
 $O_{3-6}$ are the QCD and $O_{7-10}$ are the electroweak
penguin operators.

Here we will use the QCD  factorization  approach to evaluate the hadronic matrix
elements as discussed in \cite{qcdf}. The matrix elements describing the $B_s \to \phi$
transition can be parameterized
in terms of various form factors \cite{ball} as
\bea
\langle \phi(p', \epsilon)|\bar s \gamma_\mu (1&-& \gamma_5)b|B_s(p) \rangle =-i \epsilon_\mu^*(m_{B_s}+m_\phi)A_1(q^2)\nn\\
&+ & i (p+p')_\mu (\epsilon^* \cdot q) \frac{A_2(q^2)}{m_{B_s}+m_\phi}
+ i q_\mu (\epsilon^* \cdot q) \frac{2 m_\phi}{q^2}(A_3(q^2)-A_0(q^2))\nn\\
&+ &\epsilon_{\mu\nu\rho \sigma} \epsilon^{* \nu} p^\rho p'^\sigma \frac{2 V(q^2)}{m_{B_s}+m_\phi},
\eea
where $p$ and $p'$ are the momenta of $B_s$ and $\phi$ mesons,
$q=p-p'$ is the momentum transfer, $A_{1-3}(q^2)$ and $V(q^2)$ are various form
factors describing the transition process.
Using the decay constant of $\pi^0$ meson as
\be
\langle \pi^0 (q)  |\frac{\bar u \gamma^\mu \gamma_5 u - \bar d \gamma^\mu \gamma_5 d}{\sqrt 2}
| 0 \rangle
= i \frac{f_\pi}{\sqrt 2} q^\mu \;,
\ee
one can obtain the transition amplitude for the process
\be
{\cal A}(B_s \to \phi \pi) = \frac{G_F}{ 2} f_\pi (\epsilon^* \cdot q) 2 m_\phi A_0(q^2)
\left ( V_{ub}V_{us}^* a_2 - \frac{3}{2}V_{tb} V_{ts}^*(-a_7+a_9) \right )\label{amp1}
\ee
where $\lambda_q = V_{qb} V_{qs}^*$.
The parameters $a_i$'s are related to the Wilson coefficients
$C_i$'s and the corresponding expressions can be found in
Ref. \cite{qcdf}.

The corresponding decay width is given as
\be
\Gamma(B_s \to \phi \pi) = \frac{|{\bf p}_{cm}|^3}{8 \pi m_\phi^2}
\left |\frac{ {\cal A}(B_s \to \phi \pi)}{\epsilon^* \cdot q} \right |^2\;,
\ee
where $ {\bf p}_{cm}$ is the center of mass momentum of the outgoing
particles.

Now we discuss about the CP violating observables for this process.
To obtain these observables, we can
symbolically represent the amplitude (\ref{amp1}) as \bea
 {\cal A}(\bar B_s \to \phi \pi )= (\epsilon^* \cdot q)[\lambda_u A_u -\lambda_t A_t]
=-\lambda_t A_t (\epsilon^* \cdot q)\Big[1- r~ a~ e^{-i(\pi+\beta_s+\gamma+\delta)} \Big],
\label{amp2}\eea
 where $a=|\lambda_u/\lambda_t|$,
$-\gamma$ is the weak phase of $V_{ub}$, $(\pi+\beta_s) $ is the weak
phase of $\lambda_t$,  $r=|A_u/A_t|$, and
$\delta$  is the relative strong phases between $A_t$ and $A_u$.
From the above amplitude, the direct and mixing induced  CP asymmetry
 parameters can be obtained as
\bea
A_{\phi \pi}&= &\frac{ 2 r a \sin \delta \sin(
\beta_s+\gamma)  } {1+(ra)^2+2ra \cos
\delta \cos (\beta_s+\gamma)}\nn\\
S_{\phi \pi}&= &-\frac{ 2 r a \cos \delta \sin(
\beta_s+\gamma) +(r a)^2\sin(2 \beta_s +2 \gamma) } {1+(ra)^2+2ra \cos
\delta \cos (\beta_s+\gamma)}
\;.
 \eea

For numerical evaluation,  we use the  particles masses, lifetime of $B_s $ meson
from \cite{pdg}. For the CKM elements we use the Wolfenstein parametrization
with the values of the parameters  as $\lambda= 0.2253 \pm 0.0007$,
$A=0.808_{-0.015}^{+0.022}$, $\bar \rho = 0.132_{-0.014}^{+0.022}$,
$\bar \eta=0.341 \pm 0.013$.
The parameters of QCD factorization approach  and the value of the form factor used
$A_0^{B_s \to \phi}=0.32 \pm 0.01$ are taken from \cite{cheng}.

With these inputs we obtain the branching ratio for this process as
\be
{\rm Br}(B_s \to \phi \pi) = (1.26 \pm 0.32) \times 10^{-7}\;,
\ee
which is consistent with the prediction of \cite{cheng, kim}.

The CP violating observables are found to be
\bea
S_{\phi \pi}= -0.23\;, ~~~~~~~A_{\phi \pi}= 0.1\;.
\eea
Our predicted direct CP asymmetry is lower than the prediction of \cite{cheng}.
This difference arises mainly because the sub-leading power corrections to the
color suppressed tree amplitude $a_2$ has been included in Ref. \cite{cheng},
which introduces a large strong phase.

Now we will analyze this process in the fourth generation model. In the presence
of a sequential fourth generation there will be
additional contributions due to the $t'$ quark in the
loop diagrams.
Furthermore, due to the additional fourth
generation there will be mixing between the $b'$ quark  the three
down-type quarks of the standard model and the resulting mixing
matrix will become a $4 \times 4$ matrix ($V_{CKM4})$
and the unitarity
condition becomes $\lambda_u+\lambda_c+
\lambda_t +\lambda_{t'}=0$, where
$\lambda_q=V_{qb} V_{qs}^*$. The
parametrization of this unitary matrix requires six mixing angles
and  three phases. The existence of the two extra phases provides
the possibility of extra source of CP violation \cite{hou}.
In the presence of fourth generation there will be additional contribution
both to the $B_s \to \phi \pi$ decay amplitude as well as to the
$B_s -\bar B_s$ mixing phenomenon. However, since the new physics contribution
to  $B_s - \bar B_s$ mixing amplitude due to fourth generation model
has been discussed in Ref. \cite{rm3}, we will simply quote the results
from there.

Now we will consider the additional  contribution to the decay amplitude
due to the fourth quark generation model. In this model
the new  contributions are due to the $t'$ quark in the penguin loops.
Thus, the modified Hamiltonian becomes
\bea {\cal H}_{eff}=  \frac{G_F}{\sqrt 2}\biggr[\lambda_u(C_1 O_1+C_2 O_2 )
-\lambda_t\sum_{i=3}^{10} C_i O_i-\lambda_{t'}\sum_{i=3}^{10} C_i' O_i \biggr]\;,\label{ham}
\eea
where $C_i'$'s are the effective Wilson coefficients due to $t'$ quark in the loop.
To find the new contribution due to the fourth generation effect, first we have to evaluate the new
Wilson coefficients $C_i'$. The values of these coefficients at the $M_W$ scale can be obtained
from the corresponding contribution from $t$ quark by replacing the mass of $t$ quark by $t'$
mass in the Inami Lim functions \cite{inami}. These values can then be evolved to the $m_b$ scale using the
renormalization group equation \cite{rg}. Thus, the obtained values of $ C'_{i=7-10}(m_b)$ for
two representative set of values i.e.,  $m_{t'}=400$ and 500 GeV are
as presented in Table-I.

\begin{table}[t]
\begin{center}
\caption{ Numerical values of the Wilson coefficients $C_i'$ for
$m_{t'}=400$ and 500 GeV.}
\begin{tabular}{|c|c|c|c|c|}
\hline
$t'$ mass & $C_7'$ & $C_8'$ & $C_9'$ & $C_{10}' $\\
\hline
~ $m_{t'}$=400 GeV ~& $~4.453 \times 10^{-3} $ ~&$~2.115 \times 10^{-3} $ ~&
$~-0.029~ $ &~$0.006 ~$ \\
$~m_{t'}$=500 GeV~ & $~7.311 \times 10^{-3} $ ~&$~3.199 \times 10^{-3} $ ~&
$~-0.041~ $ &~$0.009 ~$ \\
 \hline

\end{tabular}
\end{center}
\end{table}

Thus, in the presence of fourth generation model, one can obtain the transition
amplitude for $B_s \to \phi \pi$ process from Eq. (\ref{ham}),
which can be symbolically represented  as \bea
 {\cal A}(\bar B_s \to \phi \pi )&=&(\epsilon^* \cdot q)( \lambda_u A_u-\lambda_t A_t -\lambda_{t'} A_{t'})\nn\\
&=& -\lambda_t A_t(\epsilon^* \cdot q)\Big[1+ r~ a e^{i(\beta_s + \gamma - \delta)}
+r'~ b~ e^{i(\phi_s - \beta_s+\delta_1)} \Big],
\label{amp}\eea
 where $b=|\lambda_{t'}/\lambda_t|$,  $r'=|A_{t'}/A_t|$, and
$\delta_1$  is the relative strong phases between $A_{t'}$ and $A_t$.
From the above amplitude, the CP averaged branching
ratio, direct and mixing induced  CP asymmetry parameters
can be obtained as \bea
{\rm Br} &=& {\rm Br}^{\rm SM}~X\;,~~~~~~~~~~
A_{\phi \pi} =  \frac{Y}{X}\;,~~~~~~~~S_{\phi \pi}  = -\frac{Z}{X},
 \eea
with
\bea
X &=& 1+ (r a)^2 + (r'b )^2 + 2 ra \cos \delta \cos (\beta_s+\gamma) + 2 r' b \cos \delta_1 \cos(\phi_s-\beta_s)\nn\\
&+&2 r r' a b \cos(\phi_s + \gamma) \cos (\delta + \delta_1)\;,\nn\\
Y &=& 2 r a \sin \delta \sin (\beta_s+\gamma)  + 2 r' b \sin \delta_1 \sin ( \phi_s - \beta_s) + 2 r r' a b \sin(\phi_s+\gamma)
\sin (\delta+\delta_1)\;,\nn\\
Z &=& \sin 2 \theta + 2 ra \cos \delta \sin (\beta_s + \gamma + 2 \theta)-2 r' b \cos \delta_1 \sin(\phi_s -\beta_s -2 \theta)\nn\\
&+& r^2 a^2 \sin(2 \beta_s + 2 \gamma + 2 \theta) - r'^2 b^2 \sin (2 \phi_s - 2 \beta_s -2 \theta)\nn\\
&-& 2 r r' a b \cos ( \delta + \delta_1) \sin(\phi_s - 2 \beta_s - \gamma - 2 \theta). \label{cp}
\eea
In Eq. (\ref{cp}), $2 \theta$ is the additional contribution to the  $B_s - \bar B_s$ mixing phase in
the fourth generation and the expression for it can be found in Ref. \cite{rm3}.

For numerical evaluation using the values of the new Wilson coefficients
as presented in Table-I, we obtain $r \approx 7.79$, $\delta \approx 25.9^\circ$,
 $r'=3.48$ (5.03),
and $\delta_1 \approx -0.1^\circ$ ($-0.1^\circ$) for $m_{t'}=400$ (500) GeV.
For the new CKM elements $\lambda_{t'}$,
we use the allowed range of $|\lambda_{t'}|=(0.08-1.4) \times 10^{-2}$ [$(0.06-0.9) \times 10^{-2}$]
and $\phi_s = (0 \to 80)^\circ$ [$\phi_s = (0 \to 80)^\circ$] for
$m_{t'}=400$ GeV [500 GeV], extracted using the
available observables mediated through $b \to s$ transitions \cite{ref1}.
Now varying $\lambda_t'$  and $ \phi_s$ in their allowed ranges,
 we show the variation of branching ratio in the
left panel of Figure-1 and the correlation plot between
the  CP violating parameters in the right panel.
From the figure it can be seen that the branching ratio is
significantly enhanced from its SM value and large mixing-induced CP violation
($S_{\phi \pi}$) could be possible
for this decay mode in the fourth generation model. However, the
direct CP asymmetry does not deviate  much from the corresponding SM value.
It should also be noted that the branching ratio decreases slowly with the increase
of $t'$ mass. However, there is no significant $m_{t'}$ dependence of the
CP violating observables.
\begin{figure}[htb]
\includegraphics[width=8cm,height=6cm, clip]{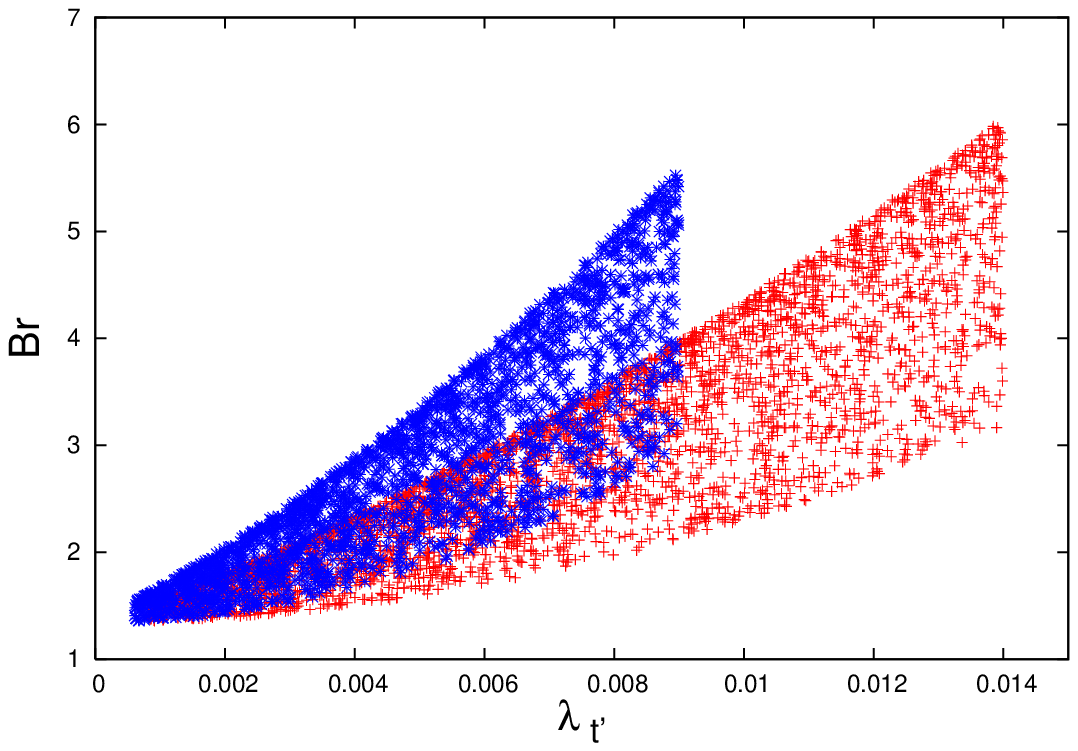}
\hspace{0.2 cm}
\includegraphics[width=8cm,height=6cm, clip]{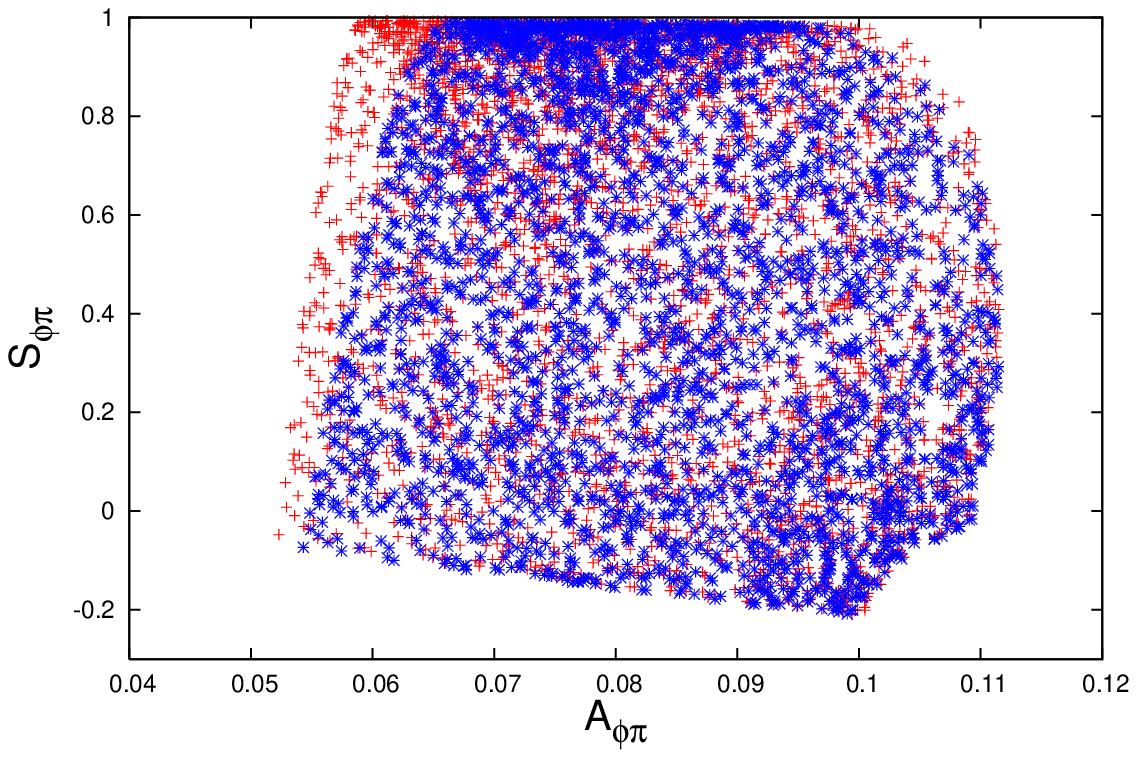}
\caption{Variation of the CP averaged Branching ratio in units of $10^{-7}$
(left panel) and the Correlation plot  between the mixing induced CP asymmetry
$S_{\phi \pi }$ and the direct CP asymmetry parameter $A_{\phi \pi}$  (right panel) for
the $B_s \to \phi \pi $  process. The red (blue) regions correspond to
$m_{t'}$=400 GeV (500 GeV) }
\end{figure}

\section{$B_s \to \phi \gamma$}

Here we will consider the decay channel $B_s \to \phi \gamma$
which is induced by the quark level transition $ \bar b \to \bar s \gamma$.
This mode is the strange counterpart of the $B \to K^* \gamma $, which is very clean
to analyze.
Compared to $B_d$ meson the new elements of $B_s$
mesons are the small mixing phase $\phi_s$ and the large width difference
$\Delta \Gamma_s$ of the $B_s$ meson. The branching ratio of this mode
is recently reported by the Belle collaboration \cite{bel}
\be
{\rm Br}(B_s \to \phi \gamma)= (57_{-15}^{+18}(\rm stat)_{-11}^{+12}(\rm syst))\times 10^{-6}\;.
\ee
In the standard model the CP averaged branching ratio of this mode is predicted to be
\cite{ball1}
\be
{\rm Br}(B_s \to \phi \gamma) = (39.4 \pm 10.7 \pm 5.3) \times 10^{-6}\;.
\ee
Although, the SM prediction seems to be consistent with the observed value, but the presence of
large experimental uncertainties makes it difficult to infer/rule out  the presence
of new physics from this mode.

The transition process $b \to s \gamma $ can be described by the dipole type effective
Hamiltonian which is given as \cite{buras}
\be
{\cal H}_{eff}= - \frac{4G_F}{\sqrt 2} \lambda_t C_7(m_b) O_7,\label{ham2}
\ee
where $C_7$ is the Wilson coefficient and $O_7$ is the electromagnetic dipole operator given as
\be
O_7= \frac{e}{32 \pi^2}F_{\mu \nu}[ m_b \bar s \sigma^{\mu \nu} (1+\gamma_5) b + m_s \bar s
\sigma^{\mu \nu} (1-\gamma_5) b ]
\ee
The expression for calculating the Wilson coefficient $C_7(\mu)$
is given in \cite{buras}.

The matrix elements of the various hadronic
currents between initial $B_s$ and the final $\phi$ meson, which are
parameterized in terms of various form factors as \cite{ball}
\bea
\langle \phi(p',\epsilon |\bar s  \sigma_{\mu \nu} q^\nu(1+\gamma_5) b | B_s \rangle & =&
i \epsilon_{\mu \nu \rho \sigma } \epsilon^{* \nu} p^\rho p'^{\sigma} 2 T_1(q^2)\nn\\
&+& T_2(q^2)[\epsilon_\mu^*(m_{B_s}^2-m_\phi^2)- (\epsilon^* \cdot q)(p+p')_\mu ]\;,
\eea
with $T_1(0)=T_2(0)$ and $q=p-p'$.
With these definition of form factors, one can obtain
the corresponding decay width as
\be
\Gamma(B_s \to \phi \gamma) = \frac{\alpha G_F^2}{32 \pi^4} |V_{tb} V_{ts}^*|^2 |C_7^{eff}|^2 m_b^2 m_{B_s}^3 |T_1(0)|^2 \left (
1- \frac{m_\phi^2}{m_{B_s}^2} \right )^3 \;. \label{br2}
\ee
Using the value of the form factor $T_1(0)=0.349\pm0.033$ \cite{ball}, $C_7(m_b)=-0.31$, and the values of
the other parameters as discussed in section II, we obtain the branching ratio as
\be
{\rm Br}(B_s \to \phi \gamma)=(39.9 \pm 12.3) \times 10^{-6}\;.
\ee
As is well known the rare radiative decays of $B$ mesons are particularly
sensitive  to the contributions from new physics.
The $V-A$ structure of the weak interactions can be tested in FCNC decays of the type
$b \to (s,d) \gamma$, since the emitted photon is predominantly left handed. The crucial
point is that the leading operator $O_7 \sim \bar s \sigma_{\mu \nu} F^{\mu \nu} b_{L(R)}$
necessitates a helicity flip on the external quark legs, which introduces a natural
hierarchy between the left and right handed component of the order $m_{d,s}/m_b$. However it
is difficult to measure the helicity of photon directly.
It was pointed out long back that the time dependent CP asymmetry is an indirect measure of
the photon helicity \cite{atwood}, since it is caused by the interference of left and right handed
helicity amplitudes. The final state in $B_s \to \phi \gamma$ is not a pure CP eigenstates. Rather in the SM
they consist of equal mixture of positive and negative eigenvalues. Thus, due to an almost
complete cancelation between positive and negative CP eigenstates, the asymmetries in
$b \to s \gamma $ is very small. They are given by $m_s/m_b$ where the quark masses
are current quark masses.

The normalized  CP asymmetry for the $B_s \to \phi \gamma$ is defined as follows \cite{zwicky}
\be
A_{CP}(B_s \to \phi \gamma)= \frac{\Gamma(\bar B_s \to \phi_s \gamma)-\Gamma(B_s \to \phi \gamma )}
{\Gamma(\bar B_s \to \phi_s \gamma)+\Gamma(B_s \to \phi \gamma)}\;,
\ee
where the left and right handed photon contributions are added incoherently i.e.,
$\Gamma(B_s \to \phi \gamma) = \Gamma(B_s \to \phi \gamma_L)+\Gamma(B_s \to \phi \gamma_R)$.
It is well known that, the neutral mesons exhibit the time dependent  CP asymmetry through mixing,
i.e., if the particle and the antiparticle decay into a common final state $f$. In $B_s
\to \phi \gamma$ this accounts to
\be
B_s \rightarrow \phi \gamma_{L(R)} \leftarrow \bar B_s
\ee

With $|q/p|=1$, the CP asymmetry assumes the following generic time dependent form
\be
A_{CP}(t)= \frac{S \sin (\Delta m_s t)-C \cos (\Delta m_s t)}{\cosh \frac{\Delta \Gamma_s t}{2}
-H\sinh{\Delta \Gamma_s t}{2}}
\ee

In terms of the left and right handed amplitudes
\be
{\cal A}_{L(R)}= {\cal A}(B_s \to \phi \gamma_{L(R)}),~~~~\bar {\cal A}_{L(R)}= {\cal A}(\bar B_s \to \phi \gamma_{L(R)}),
\ee
the form of the observables $C$, $S$ and $H$ can be found as
\bea
C=\frac{(|{\cal A}_L|^2+|{\cal A}_R|^2)-(|\bar {\cal A}_L|^2+|\bar {\cal A}_R|^2)}
{|{\cal A}_L|^2+|{\cal A}_R|^2+|\bar {\cal A}_L|^2+|\bar {\cal A}_R|^2}\;,
\nonumber\\
S=\frac{2 {\rm Im}[ \frac{q}{p} (\bar {\cal A}_L{\cal A}_L^*+\bar {\cal A}_R  {\cal A}_R^*)]}
{|{\cal A}_L|^2+|{\cal A}_R|^2+|\bar {\cal A}_L|^2+|\bar {\cal A}_R|^2}\;,\nonumber\\
H=\frac{2 {\rm Re}[ \frac{q}{p} (\bar {\cal A}_L{\cal A}_L^*+\bar {\cal A}_R  {\cal A}_R^*)]}
{|{\cal A}_L|^2+|{\cal A}_R|^2+|\bar {\cal A}_L|^2+|\bar {\cal A}_R|^2}\;.
\eea

In the standard model the leading operator $O_7$, which allows
the  $\bar B_s (B_s)$ meson to decay predominantly
into a left (right) handed photon whereas $B_s (\bar B_s)$ meson
decaying into the left (right) handed photon
suppressed by an $m_s/m_b$ chirality factor.
Due to the interference between mixing and decay in $B_s \to \phi \gamma$, a single weak
decay amplitude proportional to $\lambda_t$ is exactly canceled by the mixing phase and hence
one can obtain $S_{\phi \gamma}=0$ and $H_{\phi \gamma }= 2 m_s/m_b$ \cite{zwicky}.

The situation can be significantly modified in certain models beyond the standard model by new
terms in the decay amplitudes and also by the new contribution to
the $B_s - \bar B_s $ mixing.  In this section we
will study the effect of fourth quark generation on the various decay observables.
In the presence of fourth generation, the
Wilson coefficients $C_{7}$ will be modified due to  the new
contributions arising from the virtual $t'$ quark in the loop. Thus,
these modified coefficients can be represented as \bea
C_7^{\rm tot}(\mu) &=& C_7(\mu) + \frac{\lambda_{t'}}{\lambda_t} C_7'(\mu). \eea
The new coefficients $C_{7}'$ can be calculated at the $M_W$ scale by
replacing the $t$-quark mass by $m_{t'}$ in the loop functions.  These
coefficients then to be evolved to the $b$ scale using the
renormalization group equation as discussed in \cite{rg}. The
values of the new Wilson coefficients at the $m_b$ scale for
$m_{t'}=400$ GeV is given by $C_7'(m_b)=-0.375$.

Thus, including the new physics contribution due to fourth generation
effect the branching ratio can be obtained from Eq. (\ref{br2}) by replacing
$C_7$ by $C_7^{tot}$ and the CP violating parameters are given as
\be
S_{\phi \gamma}= \frac{m_s}{m_b} \left ( \frac{-C_7^2 \sin 2 \theta+ 2 a C_7 C_7' \sin(\phi_s-\beta_s-2 \theta)
+a^2 C_7'^2 \sin 2(\phi_s-\beta_s-\theta)}
{C_7^2 +r^2 C_7'^2 + 2 a C_7 C_7' \cos (\phi_s - \beta_s)}\right )\;,
\ee
\be
H_{\phi \gamma}= \frac{m_s}{m_b} \left ( \frac{C_7^2 \cos 2 \theta +a^2 C_7'^2 \cos
2(\phi_s -\beta_s-\theta)+2 a C_7 C_7' \cos(\phi_s-\beta_s -2 \theta) }
{C_7^2 +a^2 C_7'^2 + 2 a C_7 C_7' \cos (\phi_s - \beta_s)}\right )\;,
\ee
where $2 \theta$ is the new contribution to $B_s - \bar B_s$ mixing phase
due to fourth generation.
Now  varying $\lambda_{t'} $ between $(0.08-1.4) \times 10^{-2}$ and
$\phi_s$ between $(0-80)^\circ$ we show in Figure-2, the CP averaged branching
ratio (left panel) and the correlation plot between the CP violating observables
(right panel). From the figure it can be seen that small but nonzero CP violating
observables could be possible in the fourth generation model, while the branching ratio
still consistent with the observed value. Furthermore, in this case also the
branching ratio decreases with the increase of $t'$-mass.

\begin{figure}[htb]
\includegraphics[width=8cm,height=6cm, clip]{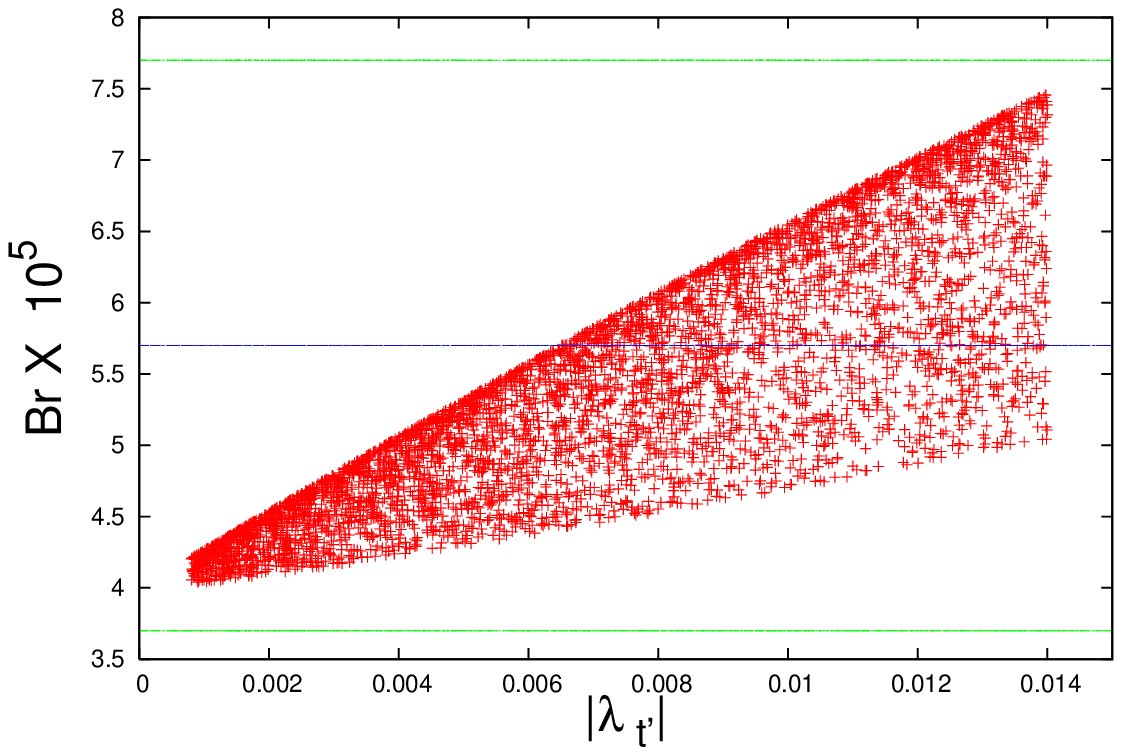}
\hspace{0.2 cm}
\includegraphics[width=8cm,height=6cm, clip]{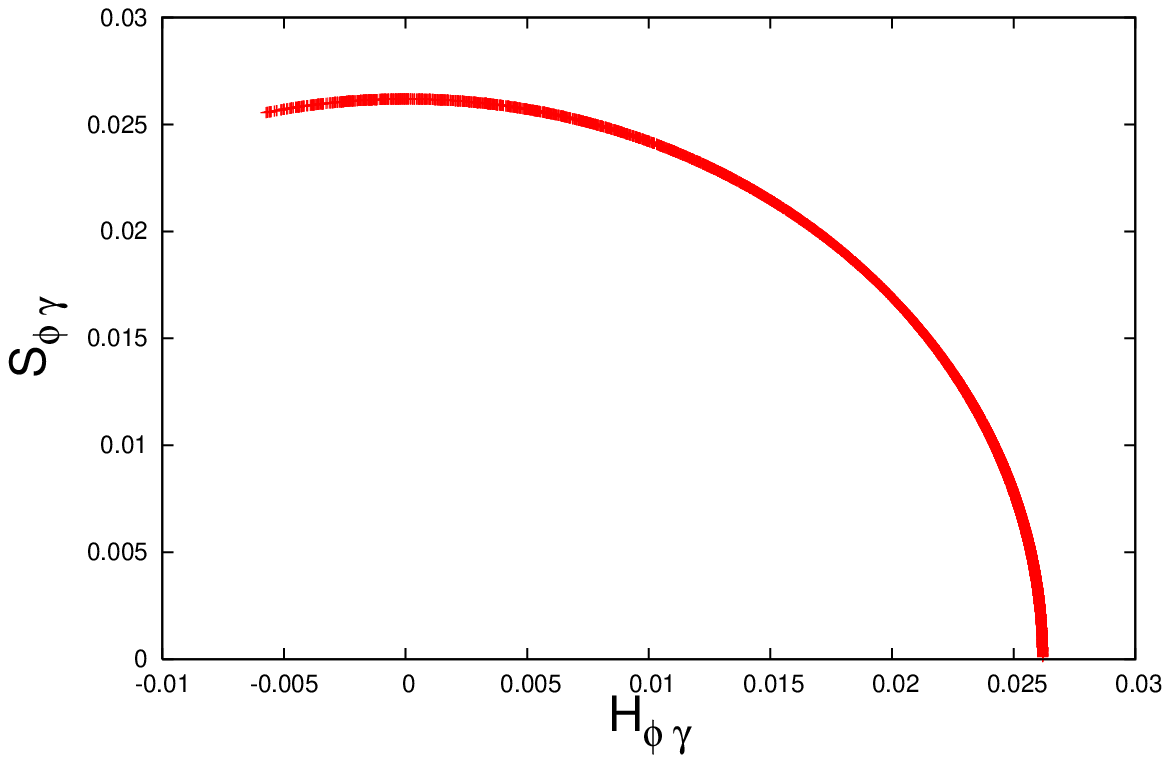}
\caption{Variation of the CP averaged Branching ratio
(left panel) and the Correlation plot  between the CP violating
observables $S_{\phi \gamma}$ and $H_{\phi \gamma}$  (right panel) for
the $B_s \to \phi \gamma $  process. The horizontal blue line on the left panel
is the central value of the measured branching ratio whereas the green lines
represent the corresponding 1-sigma range.}
\end{figure}

\section{$B_s \to \gamma \gamma $}

Now we will discuss the decay process $B_s \to \gamma \gamma $.
At the quark level this process is similar to $b \to s \gamma$. Up to the correction of
order $1/M_W^2$, the effective Hamiltonian for $b \to s \gamma \gamma $ at
scales $ \mu_b =O(m_b)$ is identical to the one for $b \to s \gamma$.
The $B_s \to \gamma \gamma $  process has been studied extensively in the SM and in
various new physics scenarios \cite{buc, gam1, gam2, gam3, gam4, gam5}.
The present experimental limit on the decay $B_s \to \gamma \gamma$ is \cite{bel}
\bea
Br(B_s \to \gamma \gamma ) \leq 8.7 \times 10^{-6}~~~~(90 \%~~C.L.).
\eea
We expect with the continuous accumulation of the experimental
data, the situation will improve and the branching ratio will be  more precise.

The effective Hamiltonian for this process is given by Eq. (\ref{ham2}). To calculate
the decay amplitude for this process one may follow the procedure discussed in Ref. \cite{gam3}.
 In order to calculate the matrix
element of Eq. (\ref{ham2}) for $B_s(p_B) \to \gamma(k_1) \gamma(k_2) $, one can work in the weak
binding approximation and assume that both the $b$ and $s$ quarks are at rest in the
$B_s$ meson and the $b$ quark carries most of the meson energy and its four velocity
can be treated as equal to that of $B_s$. Hence one may write $b$ quark momentum as
$p_b = m_b v$, where $v$ is the common four velocity of $b$ and $B_s$. Thus we have
\bea
&&p_b \cdot k_1 = m_b v \cdot k_1 = \frac{1}{2} m_b m_{B_s} = p_b \cdot k_2\nn\\
&&p_s \cdot k_1 = (p-k_1-k_2) \cdot k_1 = - \frac{m_{B_s}}{2} (m_{B_s} -m_b)= p_s \cdot k_2
\eea
The amplitude for $B_s \to \gamma \gamma$ can be computed using the following matrix elements
\bea
\langle 0| \bar s \gamma^\mu \gamma_5 b |B_s(p_B) \rangle = i f_{B_s}{ p_{B}}^\mu\nn\\
\langle 0 | \bar s \gamma_5 b |B_s \rangle = i f_{B_s} M_{B_s}
\eea

Thus, one can obtain the total amplitude for this process containing CP even and CP odd parts as
\be
{\cal A}(B_s \to \gamma \gamma) =M^+ F_{\mu \nu} F^{\mu \nu} + i M^- F_{\mu \nu} \tilde F^{\mu \nu}\;,
\ee
with
\be
M^+ = - \frac{4 \sqrt 2 \alpha G_F}{9 \pi} f_{B_s} m_{B_s} V_{tb} V_{ts}^* \left ( B m_b K(m_b^2) + \frac{3 C_7}
{8 \bar \Lambda} \right )\;,
\ee
and
\be
M^- = \frac{4 \sqrt 2 \alpha G_F}{9 \pi} f_{B_s} m_{B_s} V_{tb} V_{ts}^*
\left (\sum_q m_{B_s} A_q J(m_q^2) + m_b B L(m_b^2) + \frac{3 C_7}{8 \bar \Lambda} \right )\;,
\ee
where $\bar \Lambda =m_{B_s} -m_b$. The parameters $A_q$'s are related
to the Wilson coefficients $C_i$'s, which are evaluated at the $m_b$ scale as
\bea
A_u & = & (C_3 - C_5)N_c + (C_4-C_6)\nn\\
A_d &=& \frac{1}{4}[ (C_3-C_5)N_c + (C_4-C_6)]\nn\\
A_c &=& (C_1+C_3-C_5)N_c +C_2 +C_4 -C_6\nn\\
A_s &= & = A_b = \frac{1}{4}[(C_3+C_4-C_5)N_c +(C_3+C_4-C_6)]\nn\\
B&=& C= -\frac{1}{4}(C_6 N_c + C_5)
\eea
The functions $J(m^2)$, $K(m^2)$
 and $L(m^2)$ are defined as
\bea
J(m^2)&=& I_{11}(m^2)\nn\\
K(m^2) &=& 4 I_{11}(m^2) - I_{00}(m^2)\nn\\
L(m^2)& = & I_{00}(m^2)\;,
\eea
with
\be
I_{p q}(m^2)= \int_0^1 dx \int_0^{1-x} dy \frac{x^p y^q}{m^2- 2 k_1 \cdot k_2 x y -i \epsilon}\;.
\ee
Thus, one can obtain the decay width of $B_s \to \gamma \gamma $ is
\be
\Gamma(B_s \to \gamma \gamma)= \frac{m_{B_s}^3}{16 \pi}\left (|M^+|^2 +|M^-|^2 \right )\;.
\label{br20}
\ee

To obtain the numerical results we use the parameters as presented in section II.
Thus, we obtain the branching ratio as
\be
{\rm Br}(B_s \to \gamma \gamma)= (1.8 \pm 0.4)\times 10^{-7}\;,
\ee
which is lower than the present experimental upper bound \cite{bel}.

In the sequential fourth generation model there exist additional contribution
to $b \to s \gamma$ induced by the 4th generation up type quarks $t'$. The new Wilson
coefficients can be obtained from those of their $t$ counter parts by replacing
the mass of $t$ quark by $t'$ at the $M_W$ scale, which is then evolved to the $m_b$
scale using the renormalization group approach. As discussed in the previous section
the values of the new Wilson coefficients at the $m_b$ scale for
$m_{t'}=400$ GeV is given by $C_7'(m_b)=-0.375$.
At the scale $m_b$, the modified  Wilson coefficient of the dipole operator becomes
\be
C_{7}^{\rm tot}(m_b)= C_{7}(m_b) + \frac{V_{t'b}V_{t's}^*}{V_{tb} V_{ts}^*} C_{7}'(m_b)\;.
\ee
Now  varying $\lambda_{t'} $ between $(0.08-1.4) \times 10^{-2}$ and
$\phi_s$ between $(0-80)^\circ$ we show in Figure-3, the  branching
ratio for $B_s \to \gamma \gamma $ process.From the figure it can be  the branching ratio
can be enhanced from its SM value, but the enhancement is not so significant.

\begin{figure}[htb]
   \centerline{\epsfysize 2.5 truein \epsfbox{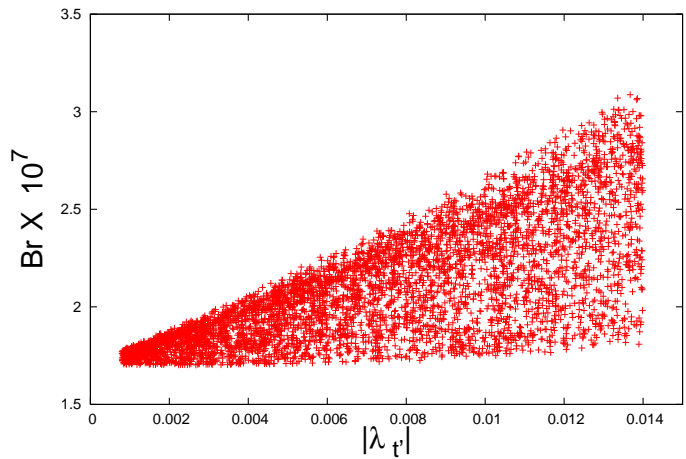}}
 \caption{Variation of the Branching ratio for the process  $B_s \to   \gamma  \gamma $
process.  }
  \end{figure}

\section{ $B_s^0 \to \mu^+ \mu^- \gamma$ process}

Now let us consider the radiative di-leptonic decay modes
$B_s \to \mu^+ \mu^- \gamma $, which are also very sensitive to the
existence of new physics beyond the SM. Due to the presence
of the photon in the final state, this decay mode is free from
helicity suppression, but it is further suppressed by a
factor of $\alpha $ with respect to the pure leptonic
$B_s \to \mu^+ \mu^-$ process. However, in spite of this
$\alpha $ suppression,  the radiative leptonic decay $B_s \to
\mu^+ \mu^- \gamma $, has comparable decay rate as that of
purely leptonic ones.

The effective Hamiltonian
describing this process $B_s \to \mu^+ \mu^-$ is \cite{rg}
\bea
\label{one}
{\cal{H}}_{eff}
&=& \frac{G_F \alpha }{\sqrt{2} \pi} V_{tb} V_{ts}^*
\Bigg[
C^{eff}_9 ~({\bar s}~ \gamma_\mu~ P_L ~b)
({\bar l}~ \gamma^\mu ~l)
+ C_{10}~({\bar s}~ \gamma_\mu~ P_L ~b)({\bar l} ~\gamma^\mu ~\gamma_5 ~l)
\nonumber \\
&&
~~~~~~~- \frac{2 C_7~ m_b}{q^2} ({\bar s}i \sigma_{\mu \nu}
q^\nu P_R ~b)
( {\bar l}~\gamma^\mu ~ l)
\Bigg]\;,\label{ham10}
\eea
where
$l$ is the short hand notation for $\mu$, $P_{L,R} = \frac{1}{2}~(1 \mp \gamma_5)$
and $q$ is the momentum
transfer. $C_i$'s are the Wilson coefficients evaluated at the $b$ quark mass
scale in NLL order with values \cite{beneke}
\be
C_7^{eff}=-0.31\;,~~C_9=4.154\;,~~C_{10}=-4.261\;.\label{wil}
\ee
The coefficient $C_9^{eff}$ has a perturbative part and a
resonance part which comes
from the long distance effects due to the conversion of the real
$c \bar c$ into the lepton pair $l^+ l^-$. Hence, $C_9^{eff}$ can be
written as
\be
C_9^{eff}=C_9+Y(s)+C_9^{res}\;,
\ee
where the function $Y(s)$ denotes the perturbative part coming
from one loop matrix elements  of the four quark operators and
is given in Ref. \cite{buras}.
The long distance resonance effect is given as \cite{res}
\bea
C_9^{res}= \frac{3 \pi}{\alpha^2}(3 C_1+C_2+3C_3+C_4+3C_5+C_6)\sum_{J/\psi,
\psi^\prime} \kappa\frac{m_{V_i} \Gamma(V_i \to l^+ l^-)}{m_{V_i}^2 -s
-i m_{V_i}\Gamma_{V_i}}\;,
\eea
where the phenomenological parameter $\kappa$ is taken to be 2.3, so as to
reproduce the correct branching ratio  $ {\cal B}(B \to J/\psi K^*
\to K^* l^+ l^-)={\cal B}(B \to J/\psi K^*){\cal B}(J/\psi \to l^+ l^-)$.
In this analysis, we will consider only
the contributions arising from two dominant resonances i.e., $J/\psi$
and $\psi^\prime$.

The matrix element for the decay $B_s \to \mu^+ \mu^- \gamma$ can
be obtained from that of the  $B_s \to \mu^+ \mu^-$ one  by attaching the photon
line to any of the charged external fermion lines. In order to
calculate the amplitude, when the photon is
radiated from the initial fermions (structure dependent (SD) part), we
need to evaluate the  matrix elements of the quark currents present
in (\ref{ham10}) between the emitted  photon and the initial $B_s$
meson. These matrix elements can be obtained by considering the
transition of a $B_s$ meson to a virtual photon with momentum $k$.
In this case the form factors depend on two variables, i.e., $k^2$ (the
photon virtuality) and the square of momentum transfer $q^2=(p_B-k)^2$.
By imposing gauge invariance, one can obtain several relations
among the form factors at $k^2=0$. These relations can be used
to reduce the number of independent form factors for the transition of
the $B_s$ meson to a real photon. Thus, the matrix elements for $B_s
\to \gamma$ transition, induced by vector, axial-vector, tensor and
pseudo-tensor currents can be parameterized as \cite{kruger}

\bea
\langle \gamma(k, \ve)|\bar s \gamma_\mu \gamma_5 b|B_s(p_B) \rangle
&=& ie \left [ \ve_\mu^* (p_B\cdot k) -(\ve^* \cdot p_B) k_\mu \right ]
\frac{F_A}{m_{B_s}}\;,\nn\\
\langle \gamma(k, \ve)|\bar s \gamma_\mu  b|B_s(p_B) \rangle
&=& e\epsilon_{\mu \nu \alpha \beta} \ve^{*\nu} p_B^\alpha~ k^\beta
\frac{F_V}{m_{B_s}}\;,\nn\\
\langle \gamma(k, \ve)|\bar s \sigma_{\mu \nu} q^\nu
\gamma_5b|B_s(p_B) \rangle
&=& e \left [ \ve_\mu^* (p_B\cdot k) -(\ve^* \cdot p_B) k_\mu \right ]
F_{TA}\;,\nn\\
\langle \gamma(k, \ve)|\bar s \sigma_{\mu \nu} q^\nu b|B_s(p_B) \rangle
&=& e \epsilon_{\mu \nu \alpha \beta} \ve^{*\nu} p_B^\alpha~ k^\beta
F_{TV}\;,
\eea
where $\varepsilon$ and $k$ are the polarization vector and the
four-momentum of photon, $p_B$ is the momentum of initial $B_s$
meson and $F_i$'s are the various form factors.

Thus, the matrix element describing the SD part takes the form
\bea
{\cal M}_{SD} &=& \frac{\alpha^{3/2}G_F}{\sqrt{2 \pi}}~
V_{tb}V_{ts}^*
 \biggr\{ \epsilon_{\mu \nu \alpha \beta}
\varepsilon^{* \nu} p_B^\alpha~ k^\beta\Big(A_1~ \bar l \gamma^\mu l
+A_2~ \bar l \gamma^\mu \gamma_5 l \Big)\nn\\
&+&
i\Big( \varepsilon_\mu^*(k \cdot p_B)-(\varepsilon^* \cdot p_B) k_\mu
\Big)\Big(B_1~ \bar l \gamma^\mu l
+B_2~ \bar l \gamma^\mu \gamma_5 l \Big)\biggr\}\;,
\label{sd}
\eea
where
\bea
A_1&=& 2 C_7 \frac{m_b}{q^2}F_{TV}+C_9 \frac{F_V}{m_{B_s}}\;,~~~
~~~~~~~~~~A_2=C_{10}\frac{F_V}
{m_{B_s}}\;,\nn\\
B_1&=& -2C_7\frac{m_b}{q^2} F_{TA}-C_9 \frac{F_A}{m_{B_s}}\;,~~~
~~~~~~~~~B_2=-C_{10} \frac{F_A}{m_{B_s}}\;.\label{ff}
\eea
The form factors $F_V$ and $F_A$ have been calculated
within the dispersion approach \cite{new1}.
The  $q^2$ dependence of the
form factors are given as \cite{kruger}
\be
F(E_\gamma)= \beta \frac{f_{B_s} m_{B_s}}{\Delta+ E_\gamma}\;,
\label{ib}
\ee
where $E_\gamma$ is the photon energy, which is related to the
momentum transfer $q^2$ as
\be
E_\gamma= \frac{m_{B_s}}{2}\left (1- \frac{q^2}{m_{B_s}^2} \right )\;.
\ee
The values of the parameters $\beta $ and $\Delta$
 are given in Table-2. The same ansatz
(\ref{ib}) has also been assumed for the form factors
$F_{TA}$ and $F_{TV}$. We use the decay constant of the $B_s$
meson, which is evaluated in lattice QCD calculation as
 $f_{B_s}=232 \pm 10$ MeV \cite{fbs}.
\begin{table}
\begin{center}
\caption{The parameters for $B_s \to \gamma$ form factors.}
\vspace*{0.3 true cm}
\begin{tabular}{|c|cc|c c|cc|cc|}
\hline
\hline
~~Parameter~~ & &~ ~$ F_V $~~ &&~ ~$ F_{TV}$~~ &&~~ $F_A$ ~~&& ~~$F_{TA}$~~ \\
\hline
$\beta ({\rm GeV}^{-1})$ && 0.28  && 0.30 && 0.26 && 0.33 \\
$\Delta ({\rm GeV})$ && 0.04  && 0.04 && 0.30 && 0.30 \\
\hline
\hline
\end{tabular}
\end{center}
\end{table}

When the photon is radiated from the outgoing lepton pairs, the
internal bremsstrahlung  (IB) part, the matrix
element is given as
\be
{\cal M}_{IB} = \frac{\alpha^{3/2}G_F}{\sqrt{ 2 \pi}}~
V_{tb}V_{ts}^*~  f_{B_s}~ m_{\mu}~ C_{10}
\biggr[ \bar l \left ( \frac{\not\!{\varepsilon}^* {\not\!{p}}_{B}}{
p_+ \cdot k}-\frac{{\not\!{p}}_{B}\not\!{\varepsilon}^* }{
p_- \cdot k} \right )\gamma_5 ~l \biggr]\;,
\ee
where $p_+$ and $p_-$ are the momenta of emitted $\mu^+$ and $\mu^-$
respectively.
Thus, the total matrix element for the $B_s \to l^+ l^- \gamma $ process
is given as
\be
{\cal M}={\cal M}_{SD}+{\cal M}_{IB}\;.
\ee
The differential decay width of the $B_s \to \mu^+
\mu^- \gamma $ process, in the rest frame of $B_s$ meson is given as
\be
\frac{d \Gamma}{d s}= \frac{G_F^2 \alpha^3}{2^{10} \pi^4}~ |V_{tb}
V_{ts}^*|^2~
m_{B_s}^3~ \Delta_1\;,\label{lp}
\ee
where
\bea
\Delta_1 &=& \frac{4}{3} m_{B_s}^2 (1- \hat s)^2 v_l \Big((\hat s+2r_l)(|A_1|^2
+|B_1|^2)+(\hat s-4 r_l)(|A_2|^2+|B_2|^2 \Big )\nn\\
&-& 64~ \frac{f_{B_s}^2}{m_{B_s}^2} \frac{r_l}{1- \hat s}~ C_{10}^2~\Big(
(4r_l-\hat s^2 -1) \ln
\frac{1+v_l}{1-v_l}+2 \hat s~ v_l\Big)\nn\\
&-& 32~r_l(1-\hat s)^2~ f_{B_s} {\rm Re}\Big( C_{10} A_1^* \Big),
\eea
with  $s=q^2$, $\hat s= s/m_{B_s}^2$, $r_l=m_{\mu}^2/m_{B_s}^2$,
$v_l=\sqrt{1- 4 m_{\mu}^2/q^2}$. The physical region of $s$ is
$4 m_{\mu}^2 \leq s \leq m_{B_s}^2 $.

The forward backward asymmetry is given as
\bea
A_{FB} &=& \frac{1}{\Delta_1} \biggr[2 m_{B_s}^2 \hat s(1-\hat s)^3 v_l^2~
{\rm Re}\Big(
A_1^* B_2+B_1^* A_2\Big)\nn\\
&+&32~ f_{B_s}~r_l (1-\hat s)^2 \ln\left (\frac{4r_l}{\hat s} \right )
{\rm Re}\Big(C_{10}B_2^*\Big)\biggr]\;.\label{fb1}
\eea

We have shown the variation of the differential  decay
distribution (\ref{lp}) (in units of $10^{-7}$,
and the forward backward asymmetry (\ref{fb1}) for $B_s \to
\mu^+ \mu^- \gamma$ in Figure-4.
\begin{figure}[htb]
\includegraphics[width=7.5cm,height=5.5cm, clip]{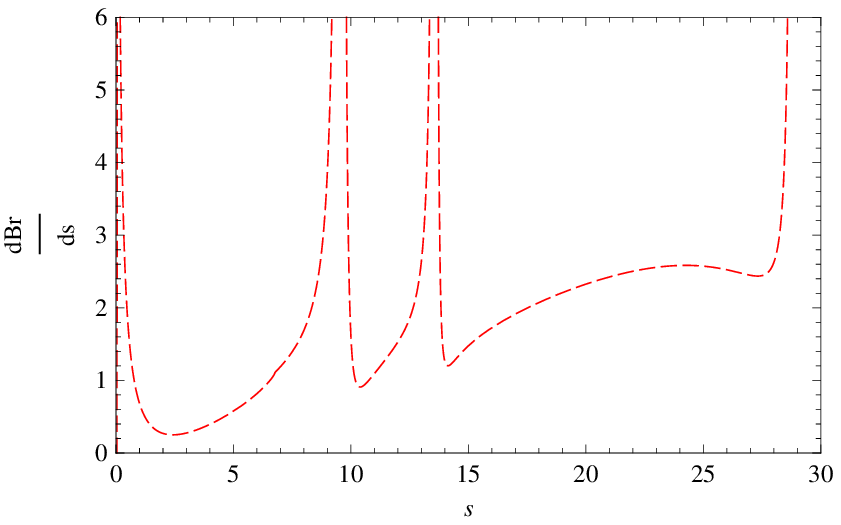}
\hspace{0.2 cm}
\includegraphics[width=7.5cm,height=5.5cm, clip]{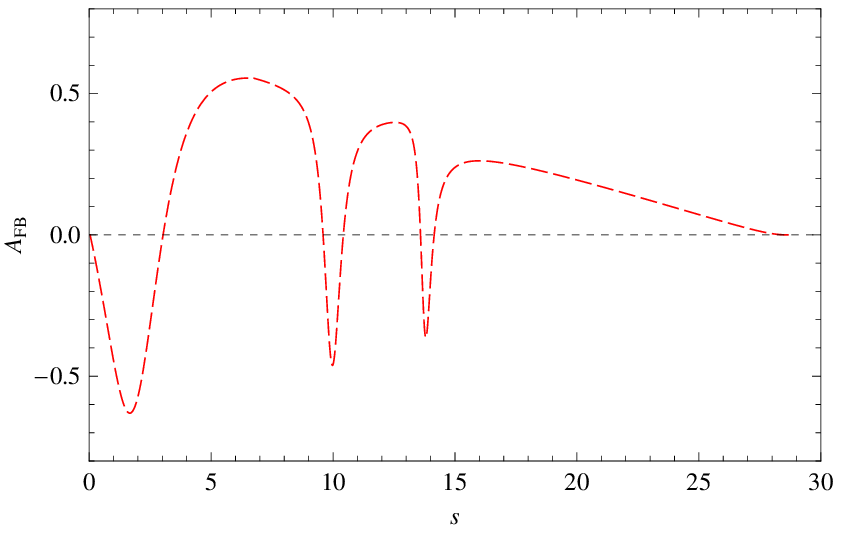}
\caption{Variation of the differential branching ratio (in units of $10^{-10}$)
(left panel) and the forward-backward  asymmetry with respect to the
momentum transfer $s$  (right panel) for
the $B_s \to \mu^+ \mu^- \gamma $  process. }
\end{figure}

As discussed earlier in the presence of fourth generation, the Wilson coefficients
$C_{7,9,10}$ will be modified due to  the new contributions arising from the
virtual $t'$ quark in the loop. Thus, these coefficients will be modified as
\bea
C_7^{\rm tot}(\mu) &=& C_7(\mu) + \frac{\lambda_{t'}}{\lambda_t} C_7'(\mu),\nn\\
C_9^{\rm tot}(\mu)&= &C_9(\mu) + \frac{\lambda_{t'}}{\lambda_t} C_9'(\mu),\nn\\
C_{10}^{\rm tot}(\mu)&= & C_{10}(\mu) + \frac{\lambda_{t'}}{\lambda_t} C_{10}'(\mu).
\eea
The new coefficients $C_{7,9,10}'$ can be
calculated at the $M_W$ scale by replacing the $t$-quark mass by $m_t'$ in the loop functions
as discussed in \cite{buras}. These coefficients then to be evolved to the $b$ scale using the
the renormalization group equation. The values of the new Wilson coefficients at the $m_b$
scale for $m_{t'}=400$ GeV is given by $C_7'(m_b)=-0.375$, $C_9'(m_b)=5.831$ and $C_{10}'=-17.358$.

Thus, one can obtain the differential branching ratio and the forward backward asymmetry
in SM4 by replacing $C_{7,9,10}$ in Eqs (\ref{lp}) and (\ref{fb1}) by $C_{7,9,10}^{\rm tot}$.
Using the values of the $\lambda_{t'}$ and $\phi_s$ for $m_{t'}=400$ GeV as
discussed earlier, the
differential branching ratio and the forward  backward asymmetry for
$B_s \to  \mu^+ \mu^- \gamma$  is presented in Figure-5, where we have
not considered the contributions from intermediate charmonium resonances. From the figure it can be seen
that the differential branching ratio of  this mode is significantly enhanced from its corresponding
SM value whereas the forward backward asymmetry is slightly reduced with respect to its SM value.
However, the zero-position of the FB asymmetry remains unchanged the fourth quark generation model.

To obtain the branching ratios it is necessary
to eliminate the background due to the resonances  $J/\psi
(\psi^\prime)$
with $J/\psi(\psi^\prime) \to \mu^+ \mu^- $. We use the
following  veto
windows to eliminate these backgrounds
\begin{eqnarray*}
&&m_{J/\psi}-0.02<
m_{\mu^+ \mu^-}<m_{J/\psi}+0.02;\nn\\
&&
 m_{\psi^\prime}-0.02<m_{\mu^+ \mu^-}<m_{\psi^\prime}+0.02 .
\end{eqnarray*}
Furthermore, it should be noted that the
$|{\cal M}_{IB}|^2$ has infrared singularity
due to the emission of soft photon. Therefore, to obtain the branching ratio,
we impose a cut on the photon energy, which will correspond to the
experimental cut imposed on the minimum energy for the detectable photon.
Requiring the photon energy to be larger than 25 MeV, i.e.,
$E_\gamma \geq \delta~ m_{B_s}/2$, which corresponds to
$s \leq m_{B_s}^2(1- \delta)$, and therefore, we set
the cut $\delta \geq 0.01 $.
Thus, with the above defined veto windows and the infrared cutoff parameter,
the total branching ratio for $B_s \to \mu^+ \mu^- \gamma$ process is found to be
\bea
{\rm Br}(B_s \to \mu^+ \mu^- \gamma) & =& 5.6 \times 10^{-9}\;,~~~~~~{(\rm SM)}\nonumber\\
& < & 2.2 \times 10^{-8}\;,~~~~~~~{\rm (for~ m_{t'}=400~ GeV)}\;.
\eea
The above branching ratio is comparable with that of the corresponding pure-leptonic
process, $B_s \to \mu^+ \mu^-$, whose predicted branching ratio \cite{ref1} for
$m_{t'}=400 $ GeV  is
\bea
{\rm Br}(B_s \to \mu^+ \mu^- )
& < & 0.8 \times 10^{-8}\;.
\eea
The LHCb \cite{lcb} has searched for this process and set the upper limit as
${\rm Br}(B_s \to \mu^+ \mu^-)
 <  1.2~ (1.5)\times 10^{-8} $ at $90 \%$ ($95 \%$) CL. Therefore, the
$B_s \to \mu^+ \mu^- \gamma$ decay channel could also be accessible there
and hopefully it will be observed soon.

\begin{figure}[htb]
\includegraphics[width=8cm,height=6cm, clip]{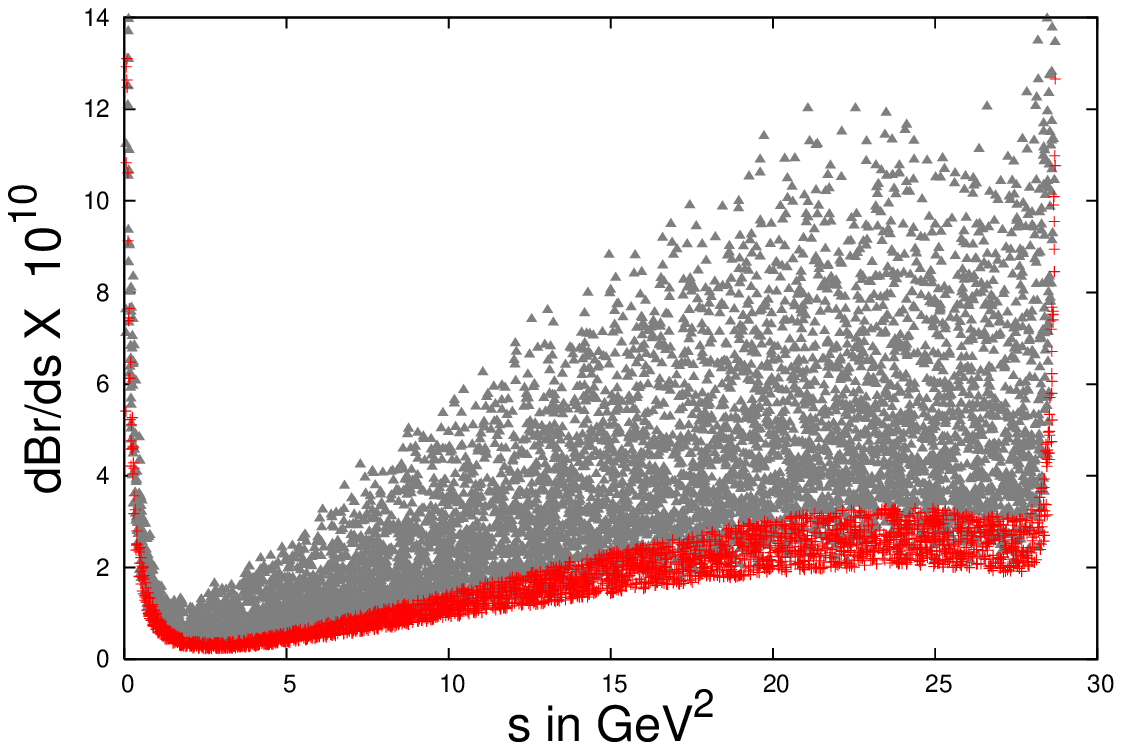}
\hspace{0.2 cm}
\includegraphics[width=8cm,height=6cm, clip]{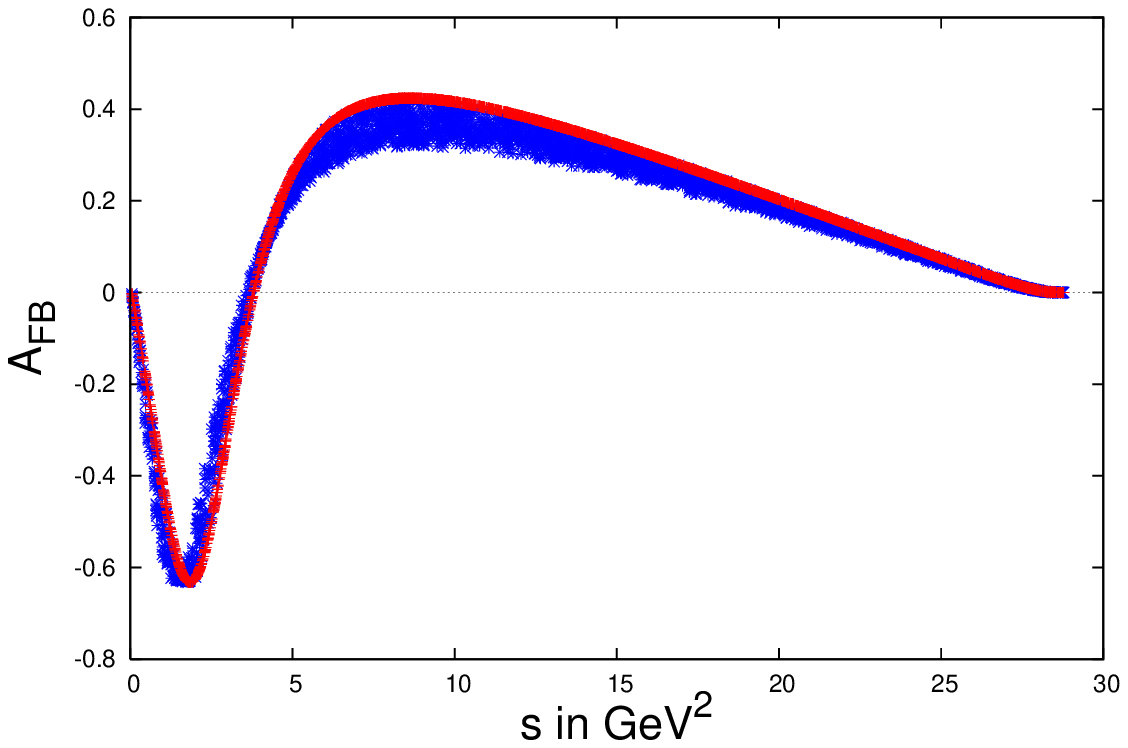}
\caption{Variation of the differential branching ratio
(left panel) and the forward-backward asymmetry with respect to the
momentum transfer $s$  (right panel) for
the $B_s \to \mu^+ \mu^- \gamma $  process, in fourth quark generation
model (red regions) whereas the corresponding SM values are shown
in blue regions.}
\end{figure}
\section{Conclusion}
In this paper we have studied some rare decays of the $B_s $ meson in the fourth
quark generation model. The large production of $B_s$ mesons at the LHC opens up the possibility
to study $B_s$ meson with high statistical precision.
The decay modes considered here are $B_s \to \phi \pi$, $B_s \to \phi
\gamma$, $B_s \to \gamma \gamma$ and $B_s \to \mu^+ \mu^- \gamma$, which are
highly suppressed in the SM as they occurred only through one-loop  diagrams.
Therefore, they provide an ideal testing ground to look for new physics.
The fourth generation model is a very simple extension of the SM with three generations and it
can easily accommodate the observed anomalies in the $B$ and $B_s$ CP violation
parameters for $m_{t'}$ in the range of (400-600) GeV.
 We found that in the fourth generation model
the branching ratios for these processes enhanced from their corresponding
SM values.  However,  the mixing-induced CP asymmetry of in $B_s \to \phi \pi$ process
enhanced significantly from its SM value. The CP violating observables
in $B_s \to \phi \gamma$ are found to be small but nonzero.
Some of these branching ratios are within the reach of LHCb experiments, hence the observation
of these modes will provide us an indirect evidence for the existence of fourth quark
generation.

{\bf Acknowledgments}

RM would like to thank Council of Scientific and Industrial Research,
Government of India, for financial support through Grant No.
03(1190)-11/EMR-II.

\end{document}